\documentclass[12pt]{article}
\usepackage{amsmath}
\usepackage{amsfonts}
\usepackage{graphics}
\usepackage{setspace}
\usepackage{rotating}

\begin{document}

\begin{titlepage}
\title{\vskip -70pt
\begin{flushright}
{\normalsize \ }\\
\end{flushright}
\vskip 20pt
{\bf Twisted Vortices in a Gauge Field Theory }
}
\vspace{1cm}
\author{
{M. L\"ubcke$^{\, 1}$}
\hspace{.15cm}
{S M. Nasir$^{\, 2}$}
\hspace{.15cm}
{A. Niemi$^{\, 1}$}
\hspace{.15cm}
{K. Torokoff$^{\, 1}$}
\\ 
{ \sl Department of Theoretical Physics, Uppsala University }
\\
{ \sl Box 803, S-75 108 Uppsala, Sweden}}

\maketitle
\thispagestyle{empty}
\vspace{1cm}
\begin{abstract}
\noindent
We inspect a particular gauge field theory model that describes 
the properties of a variety of physical systems, including
a charge neutral two-component plasma, a Gross-Pitaevskii 
functional of two charged Cooper pair condensates, and
a limiting case of the bosonic sector in the Salam-Weinberg 
model. It has been argued that this field theory model 
also admits stable knot-like solitons. Here we produce 
numerical evidence in support for the existence 
of these solitons, by considering 
stable axis-symmetric solutions that can be thought of as straight 
twisted vortex lines clamped at the two ends. We compute the 
energy of these solutions as a function of the amount of twist 
per unit length. The result can be described in terms of a 
energy spectral function. We find that this spectral function 
acquires a minimum which corresponds to a nontrivial twist
per unit length, strongly suggesting that the model indeed supports 
stable toroidal solitons.   
\end{abstract}

\begin{flushleft}
\rule{5.1 in}{.007 in} \\
{\small \tt Martin.Lubcke@teorfys.uu.se, Nasir@teorfys.uu.se, 
Niemi@teorfys.uu.se, Kristel.Torokoff@teorfys.uu.se} \\
$^1$ \hskip 0.5cm supported by NFR Grant F-AA/FU 06821-308 \\
$^2$ \hskip 0.5cm supported by G\"{o}ran Gustafssons Stiftelse UU/KTH
\end{flushleft}

\end{titlepage}

\noindent
Recently, a gauge field theory model with two charged bosons
has been conceived, to describe a two-component plasma
of negatively and positively charged particles \cite{fn3}.
But this model also appears to describe 
a large variety of other physical phenomena, including
a Gross-Pitaevskii functional
of two band superconductivity \cite{egor} and 
the bosonic sector in the Salam-Weinberg model in the limit where 
the Weinberg angle $\theta_{W}=0$ \cite{sami}.
In \cite{fn3} (see also \cite{sami}) it has been 
proposed that the model supports stable, self-confining and 
knot-like solitonic configurations. This would be somewhat
remarkable, since it would {\it contrast}
some of the widely held views in plasma physics that such 
configurations of plasma can not exist in general. This 
is due to a simple application of the Shafranov virial 
theorem which states that a static configuration of plasma 
in isolation is dissipative \cite{fr}. The proposed model, 
however, escapes this no-go theorem by incorporating non-linear 
interactions which are not accounted for by mean field 
variables such as the pressure \cite{lisa}.

\noindent
The soliton solutions in the gauge model that we shall inspect
can be viewed as a bundled filaments of twisted magnetic flux 
lines. The twisting is governed by a certain topological quantity, 
the Hopf invariant. Nontriviality of the Hopf invariant 
ensures that the flux lines are knotted, or linked.
Numerical simulations, in the absence of effective analytical 
tools, seem so far to be the best way to help explore the nature 
of the soliton solutions. But even then their intricate topology 
makes full three dimensional simulations a daunting task. 
In this letter we present and analyse a tractable, yet challenging, 
simulation of the model, where the magnetic flux lines are twisted 
in an axis-symmetric manner. Such configuration of solutions can 
then be viewed as straight but twisted vortex tubes. 

\noindent
As such, the stable vortices in the model we study can be applied to
study a number of interesting physics. They may relate to the 
coronal loops on the solar photosphere \cite{fn3}, to Meissner effect
in two-band superconductors \cite{egor} or to higher 
energy topological configurations in the weak sector of the 
standard model \cite{sami}, \cite{cho}.
Our study will then serve as a test bed to understand knot 
solitons in general. Serious, three dimensional searches for 
knotted structures in field theory models are attempted only 
recently. One prototype model, initiating these searches 
\cite{f}, \cite{fn1} (see also \cite{gh, bs, jarmo}), is a Skyrme like 
$O(3)$ non-linear sigma model. This model 
could be envisaged as describing the infrared phase of the 
pure $SU(2)$ Yang-Mills theory, with glueballs represented by 
the knotted flux tubes of gluon \cite{fn2}. 

\noindent
We start from the classical kinetic theory model of a 
two-component plasma of 
electromagnetically interacting electrons and ions, given 
by the non-relativistic action \cite{fn3},
\begin{eqnarray}
S & = & \int d^{4}x\left[ i\hbar \Psi _{e}^{*}\left( \partial _{t}+
\frac{ieA_{t}}{\hbar c}\right) \Psi _{e}+i\hbar \Psi _{i}^{*}\left( 
\partial _{t}-\frac{ieA_{t}}{\hbar c}\right) \Psi _{i} \right. 
\\ \nonumber &  & \left. -\frac{\hbar ^{2}}{2m}\left| \left( 
\partial _{k}+\frac{ieA_{k}}{\hbar c}\right) \Psi _{e}\right|^{2}
-\frac{\hbar ^{2}}{2M}\left| \left( \partial _{k}-\frac{ieA_{k}}
{\hbar c}\right) \Psi _{i}\right| ^{2}-\frac{1}{4}F^{2}_{\mu \nu }\right] .
\end{eqnarray}
Here, \( \Psi _{e} \) and \( \Psi _{i} \) are the two complex 
non-relativistic, macroscopic Hartree wave functions
describing the electrons ({\it e}) and 
ions ({\it i}) with their respective masses \( m \) and \( M \). 
Numerically, with deuterons we have \( {\displaystyle{\alpha 
=\frac{m}{M}= \frac{1}{3670}}} \). The electron and ion densities 
are, respectively, given by $ \Psi_{e}^{*}\Psi_{e}$ and 
$\Psi_{i}^{*}\Psi_{i}$, and their total integrals over the 
three-space give the total electron number $N_{e}$ and the total 
ion number $N_{i}$. Charge 
neutrality requires $N_{e}=N_{i}$.

\noindent
The ensuing static Hamiltonian in the Coulomb gauge is, 
\begin{eqnarray}\label{e'}
H & = & \int d^{3}{\bf{x}}
\left[ 
\frac{\hbar ^{2}}{2m}\left| \left( \partial _{k}
+\frac{ieA_{k}}{\hbar c}\right) \Psi _{e}\right|^{2}
+\frac{\hbar ^{2}}{2M}\left| \left( \partial _{k}
-\frac{ieA_{k}}{\hbar c}\right) \Psi _{i}\right|^{2}
+\frac{1}{2}{\bf{B}}^{2}\right] ,
\end{eqnarray}
where ${\bf{B}}$ is the magnetic field. We note
the similarity of the above with the Hamiltonian that
describes the the bosonic sector of 
the Salam-Weinberg model at $\theta_{W}=0$: At 
this prescribed value of the Weinberg angle the masses of $W^{\pm}$ 
and $Z$ boson become infinite and they, therefore, 
decouple from the theory. Now, assigning the hypercharge 
matrix of the Higgs doublet to be proportional to the third 
Pauli matrix, $\tau_{3}$, the static Hamiltonian of the 
bosonic sector of the Salam-Weinberg model is, 
\begin{eqnarray}
H_{SW} & = & \int d^{3}{\bf{x}}
\left[ 
\frac{1}{2}\left| \left( \partial _{k}- ie A_{k}\tau_{3}
\right) \Phi \right| ^{2} 
+\frac{\mu^{2}}{2} \Phi^{\dagger}\Phi -\frac{\lambda}{4}
(\Phi^{\dagger}\Phi)^2
+\frac{1}{2}{\bf{B}}^{2}\right] ,
\end{eqnarray}
where the Higgs doublet is given by,
\begin{equation}\nonumber 
\Phi = \left( \begin{array}{c}
 \phi^{+} \\ \phi^{-} 
\end{array} \right). 
\end{equation}
In the limit of weak self-couplings between the Higgs 
fields $H_{SW}$ is notably similar to the Hamiltonian in 
Eqn.(\ref{e'}) with the obvious identification 
$\phi^{+,-}\equiv \Psi_{e,i}$.
 
\noindent
An effective static energy functional of plasma can be obtained from 
Eqn.(\ref{e'}) in a self-consistent gradient expansion 
by keeping terms with at most fourth order in the 
derivatives of the variables. 
In order to describe the ensuing tubular field configurations 
appearing in the model, it is natural to introduce a new 
set of variables \cite{fn3},
\begin{equation}
\left( \Psi _{e},\Psi _{i}\right) =\rho \left( \cos 
\alpha \cdot \sin \frac{\theta }{2}e^{i\varphi },\sin 
\alpha \cdot \cos \frac{\theta }{2}e^{i\chi }\right) ,
\end{equation}
where \( \alpha  \) is a parameter expressed in terms 
of the reduced mass, $\mu$, through the relation \( 
\mu =m \sin ^{2}\alpha =M \cos ^{2}\alpha  \),
and \( \rho ^{2} \) is related to the plasma density, 
and \( \theta  \) is a shape
function measuring roughly the distance from the center 
of the configuration,
and \( \varphi  \) and \( \chi  \) are the toroidal and 
poloidal coordinates in $ {\mathrm{{\bf{R}^{3}}}}$.
By defining a three component unit vector, \textbf{
\[
\overrightarrow{\mathbf{n}}=\left( \cos \left( \chi 
+\varphi \right) \sin \theta ,\; \sin \left( \chi 
+\varphi \right) \sin \theta ,\; \cos \theta \right), \]
} it can be shown that the static energy is \cite{fn3},
\begin{eqnarray}\label{e4}
E & = & \int d^{3}{\bf{x}}\left[ C_{1}\left( 
\partial _{k}\rho \right) ^{2}+C_{2}\rho ^{2}\left| 
\partial _{k}\overrightarrow{\mathbf{n}}\right|^{2}
+C_{3}\left( \overrightarrow{\mathbf{n}}\cdot 
\partial _{i}\overrightarrow{\mathbf{n}}\times 
\partial _{j}\overrightarrow{\mathbf{n}}\right)^{2} 
\right. \\ \nonumber
&  & \left. +\; C_{4}\rho ^{4}\left( \cos 2\alpha 
-\cos \theta \right) ^{2}\right] ,
\end{eqnarray}
where \( {\displaystyle{C_{1}=\frac{\hbar ^{2}}
{8\mu }\sin ^{2}2\alpha}}  \), \( {\displaystyle{C_{2}
=\frac{C_{1}}{4}=\frac{\hbar ^{2}}{32\mu }\sin^{2}
2\alpha }} \), \( {\displaystyle{C_{3}=\frac{\hbar^{2}
c^{2}}{8e^{2}}}} \), and \({\displaystyle{ C_{4}
=\frac{g}{4} }}\). The effective coupling \( g \) 
describes the remnant of the Coulomb interaction 
in the plasma, in the limit of short Debye screening length. 
At this point, it is of interest to 
compare the above energy density with that of \cite{fn1}. There,
the energy density consists of the two middle terms in
the above expression, for a constant $\rho$. Indeed, 
the presence of a nontrivial coupling between 
$\rho$ and $\overrightarrow{\mathbf{n}}$ in the above
expression for energy density is especially
noticeable.

\noindent
In order to have finite energy configurations, asymptotically 
at large distances \textbf{\( \overrightarrow{\mathbf{n}} \)}
must go to a constant value with \( n_{3}=\cos 2\alpha 
\), and also \( \rho =\rho _{0} \) asymptotically at 
large distances. Here, 
\( \rho _{0} \) is a constant valued characteristic 
plasma parameter related to the plasma density
at the bulk. For example on the solar photosphere 
\( \rho _{0} \) is of order of magnitude 
\( 10^{15}/{\rm{m}}^{3} \). 
The unit vector $\overrightarrow{\mathbf{n}}$, when 
combined with the boundary conditions, describes a map 
from the compactified $\mathbf{R}^3$ to the target 
$\mathbf{S}^2$.  Under this map the pre-image of a point 
on the target is generically a circle, knotted or 
linked, and such circle is a constituent element of 
the magnetic field lines in the plasma. Any two pre-image 
circles are linked with their Gauss linking number 
given by the topologically invariant, integer valued Hopf
number \( H \),  
\begin{equation}
H=\frac{1}{4\pi ^{2}}\int d^{3}{\bf{x}}\;  
\overrightarrow{\mathbf{A}}\cdot \overrightarrow{\mathbf{B}}.
\end{equation}
Stable finite energy knotted and linked 
soliton solutions are
classified by the Hopf invariant $H$. A non-trivial 
question of interest is to answer as to for which 
Hopf numbers the solutions are actually knotted, 
not merely linked.  

\noindent
The equations of motion arising from varying Eqn.(\ref{e4}) 
depend on two parameters \( \rho _{0} \) and \( g \). 
However, by re-scaling \( \rho \rightarrow \rho _{0}
\widetilde{\rho } \) and \( x\rightarrow x_{0}\widetilde{x} 
\), where \( \widetilde{\rho } \) and \( \widetilde{x} 
\) are both dimensionless quantities, the equations of motion
can be recast to make dependent only on \( g \). Henceforth, 
all the expressions are written in terms of the 
dimensionless variables \( \widetilde{\rho } \) and
\( \widetilde{x} \) and we continue to denote them 
as \( \rho  \) and \( x \), respectively. The parameter 
\( x_{0} \) has the dimension of length and has the 
expression \( x_{0}^{2}=\displaystyle{\frac{C_{3}}
{C_{1}\rho _{0}^{2}}} \). 

\noindent
We are interested in the axially symmetric solutions 
to obtain straight twisted vortices. The vector field 
generating such axis-symmetric twist is 
\[
{\displaystyle{
V=(\frac{1}{k} \partial_{\phi}-\frac{1}{a}\partial_{z})}}
\] 
and the Lie derivative of the field variables with respect 
to $V$ must be zero. The ansatz for the fields, satisfying 
the previous condition, are: \( (\chi +\varphi )=
kz+ar \), \( \rho =\rho \left( r\right)  \),
and \( \theta =\theta \left( r\right)  \). Here, \( 
k \) is a real number, and \( a \) denotes the twist 
per unit length. Similar ansatz were used in \cite{mns}. 
We consider the tube to be clamped at its two ends 
and the length of the tube is $L$. The Hopf invariant 
now becomes \( H={\displaystyle{\frac{kaL}{2\pi }}} \). 
Notice that, had we taken the tube length to be infinite, 
the Hopf number would have become infinite. It might 
be tempting to consider a straight tube with the 
topology of a torus, but this does not work:  
The toroidal topology implies that the fields are 
periodic in $z$ with period $L$, which in turn 
implies that $kL$ is an integer multiple of $2\pi$. 
One would then have $a$ to be a rational number only.
This is why we exclude henceforth a straight tube 
with only toroidal topology.

\noindent
The energy functional Eqn.(\ref{e4}) in the axially 
symmetric ansatz reads,
\begin{eqnarray}\label{en2}
\mathcal{E} & = & \lambda \int rdr\left[ \left( 
\partial _{r}\rho \right) ^{2}+\frac{1}{4}\rho^{2}
\left( \left( \partial _{r}\theta \right)^{2}
+\sin ^{2}\theta \left( \frac{k^{2}}{r^{2}}
+a^{2}\right) \right)\; \; +\right. \\ \nonumber
&  & \left. \sin ^{2}\theta \left( \partial _{r}
\theta \right) ^{2}\left( \frac{k^{2}}{r^{2}}+
a^{2}\right) +C\rho ^{4}\left( \cos 2\alpha -
\cos \theta \right) ^{2}\right] ,
\end{eqnarray}
where the prefactor \( {\displaystyle{\lambda =
\frac{\sqrt{A_{1}A_{3}}\rho _{0}(2\pi )^{2}H}{ak} }}\), 
and the Coulomb coefficient \( {\displaystyle{C=
\frac{C_{4}C_{3}}{C^{2}_{1}} }} \). Extraction of 
the parameter dependence of the field variables
of the above energy functional
is particularly revealing, as we will see shortly.

\noindent
The numerical solution to the Euler equations of motion 
arising from (\ref{en2}) are obtained by seeking the fixed points
of the following system of equations: 
\begin{equation}
\dot{\rho }=-\frac{\delta E}{\delta \rho }, 
\end{equation}
\begin{equation}
\dot{\theta } = -\frac{\delta E}{\delta \theta }.
\end{equation}
The simulations are run on a lattice of finite size.
At one boundary end, we take \( \rho=1 \),
and \( \theta = 2\alpha  \).
At the other end, which is the origin, we also 
need the boundary values  of \( \rho \) and \( \theta \). 
For these values to be fixed, first
note that the equations of motions are invariant under 
the parity transformation \( r\rightarrow -r \) and, 
therefore, we are free to choose independently \( 
\theta \) and \( \rho \) to be either odd or even function 
in $r$. We have \( \theta (0)=0 \) corresponding to 
choosing $\theta$ to be odd. As we want \( \rho \) to be 
non-vanishing at the origin, we take
 \( \rho \) to be even. It should be mentioned that 
\( \rho \) could have been chosen equally well to be an 
odd function, but we would like charge densities to be 
non-vanishing at the origin. With the boundary conditions 
so chosen and for technical reasons in order to facilitate 
the simulations, the range of the lattice is also 
extended to the negative values of $r$. Next, we choose 
initial profiles for $\rho$ and $\theta$  matching with 
the boundary conditions. Finally, the equations are 
solved for fixed \( k=1 \), since higher \( k\) would 
corresponds to the configurations with higher energies and are,
therefore, excluded from our simulations. In the simulations
we have performed the Coulomb 
term \( C \) is chosen to be of small value, \( 0.1 \), 
\( 1.0 \), and \( 10.0 \), and the twist parameter \( 
a \) is made to lie in the range \( \left[ 0.2,2.0\right]  \). 

\noindent
In Figure 1. we have drawn plots for the energy per unit Hopf number, 
${\mathcal{E}}/H$, against the twist per unit length $a$. 
Each point on the plot corresponds to a solution of the 
equations of motion for a given $a$. These energy plots for 
different Coulomb couplings can be described by spectral 
functions $f(a,C)$. As visible from the plots, these spectral
functions all have the following features in common: 
For each $C$ the spectral functions are smooth, positive, and 
strictly convex with a nontrivial minimum. That for a given $C$ 
the spectral function $f(a,C)$ is a positive convex function 
of the twist $a$ could be seen apriori from the form of the 
energy Eqn.(\ref{en2}). As  both for $a\rightarrow 0$ and 
$\infty$, ${\mathcal{E}}/H$  diverges, given that the solutions 
are smooth. However, what is remarkable is the form of the graph 
of the function. The unique minimum point of the graph, 
occurring at $a_{C}$, represents the true stable solution 
that an axis-symmetric vortex tube with a given number of 
twist would settle to. Too many, or too few, twists per unit length 
in the tube to begin with would result in instability. 
It is to note that as $C$ varies so does $a_{C}$, but little. 
We have performed the numerical simulations for a number of
representative values of the parameters, quite far away from
realistic values applicable {\it e.g.} to coronal loops on the
solar photosphere. It would certainly be of interest to extrapolate 
our calculations for the physically interesting values of $C$.
Unfortunately, in this case 
the various numerical parameters involved deviate
from each other by several orders of magnitude. As a consequence
we find numerical intractability as a
hindrance for achieving this goal, and postpone it to future 
publications.

\noindent
The plots for the energy and ion densities have also some 
interesting features, as described in Figs. 2 and 3. Namely the 
peak of the energy density plot lies, somewhat counter-intuitively, 
slightly off the center. The reason for this can be traced to 
the twisting of the field lines. By looking at the energy 
density plot, one can furhtermore estimate the thickness of 
the vortex tube and, on the other hand, by reading off the 
minimum point of the spectral curve one can estimate its 
length. We find that the ratio between the length and the 
thickness turns out to be 2.5. This suggest that for a 
would-be toroidal configuration the energy density becomes 
lumped at the center of the toroidal structure in analogy
with the model \cite{f}; see \cite{bs, jarmo}.

\noindent
The total number of ion and electron numbers, respectively, 
$N_{e}$ and $N_{i}$, are tabulated in Table 1 for different 
values of $a$ and $C$. Clearly, we do not obtain $N_{e} = 
N_{i}$ implying the vortices carry charge, but this we consider
to be a finite size-effect as the simulations are run on 
a finite lattice. One can, however, conclude from the table 
that the heavier ions are concentrated more towards the 
center of the tube and the lighter electrons
are spread out more to the bulk.

\noindent
To conclude, we have presented numerical evidence that the 
gauge field theory model of plasma \cite{fn3} does admit 
stable knotted or linked soliton solutions. We have searched 
for a particular kind of soliton in the shape of straightened 
twisted tube. The length of the tube is clamped at a fixed 
length and the stability of the solution requires the tube 
to be twisted by a certain amount per unit length. It would be 
of interest to extend our analysis to the toroidal case, 
where the addition of curvature demands a full three dimensional 
simulation. And it would certainly be desirable to run simulations 
in the physically interesting regime of the parameter space
in order to understand for example the origin of the coronal loops
on the solar photosphere, or knotted configurations in the
Salam-Weinberg model. At the moment this is hampered by technical
reasons as the various parametes involved deviate from
each other by several orders of magnitude, in particular in the case of
solar surface. This unfortunately undermines the numerical 
stability in our present simulations. Besides finding 
plausible applications in areas of plasma and condensed matter 
physics, our study appeals also directly to the understanding 
of knot solitons in gauge field theories in general. Indeed, a 
highly interesting question would be whether 
the weak sector of the standard model admits knot solitons
\cite{sami, cho}. \\


\newpage
\begin{table}[ctb]
\begin{tabular}{||l|l|l|l|l|l|l|l|l||}\hline
&
\multicolumn{2}{c|}{\( C_{4}=0.1 \)}&
\multicolumn{2}{c|}{\( C_{4}=1.0 \)}&
\multicolumn{2}{c|}{\( C_{4}=10.0 \)}
\\\cline{2-7}
\( a \) & \( N_{i}\) & \( N_{e}\)& \( N_{i}\)& \( N_{e} \) &
\( N_{i} \) & \( N_{e} \)\\ \hline

\( 0.2 \)& 5632& 784& 5554& 641 & 1448& 633\\

\( 0.3 \)& 3334& 774 & 2753& 716& 1384& 623\\

\( 0.4 \)& 2057& 751& 1821& 720& 1061& 628\\

\( 0.5 \)& 1325& 722&1250&706&1019&622\\

\( 0.6 \)& 910& 693&888 &685&782& 628\\

\( 0.7 \)&

 667& 666& 660& 661& 619& 624\\
\( 0.8 \)&

515& 641& 512& 638 & 494& 613\\
\( 1.0 \)&

342& 596&342&595&338&583\\
\( 1.2 \)&

250&557& 250&557&249&551\\
\( 1.4 \)&

194&524&194&524&193&520\\
\( 1.6 \)&

157&495&157&494&157&492\\
\( 1.8 \)&

131&469&131&469&131&467\\
\( 2.0 \)&

112&446&112&446&112&445\\ \hline
\end{tabular}
\caption{ Total number of ions, $N_{i}$, and electrons, 
$N_{e}$, for different values of the twist per unit length, 
$a$, and the Coulomb coupling, $C$.
}
\end{table} 

\begin{figure}[htb]
{\centering \resizebox*{9cm}{!}{\rotatebox{0}{\includegraphics
{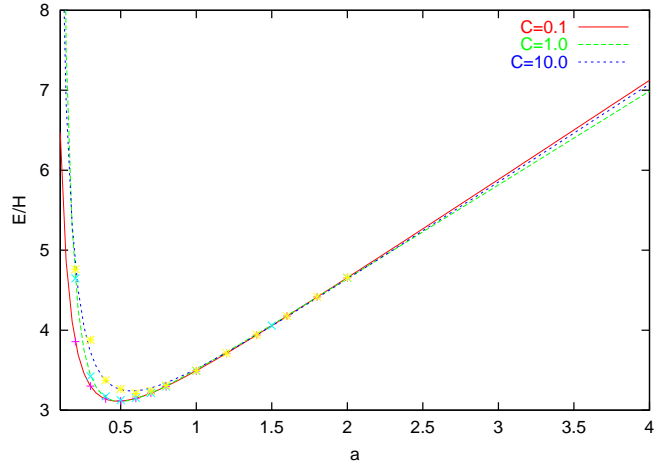}}} \par}

\caption{\label{Fig:Ea}The total Energy per unit Hopf 
number as a function of $a$, for different
values of $C$.  }
\end{figure} 

\begin{figure}[htb]
{\centering \includegraphics{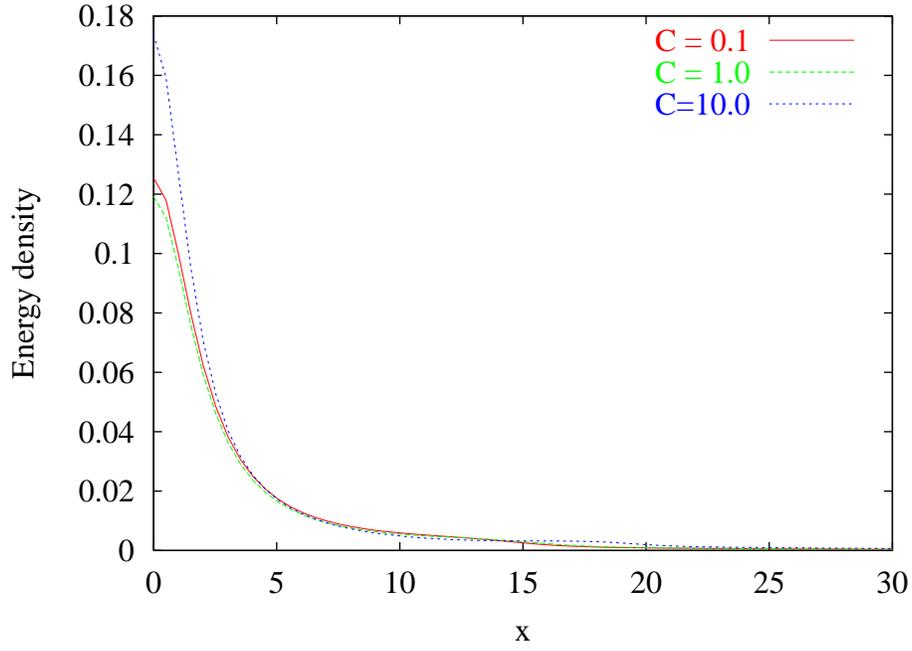} \par}

\caption{\label{Fig:energy} The energy density versus distance for different
values of $C$. }
\end{figure}

\begin{figure}[htb]
{\centering \includegraphics{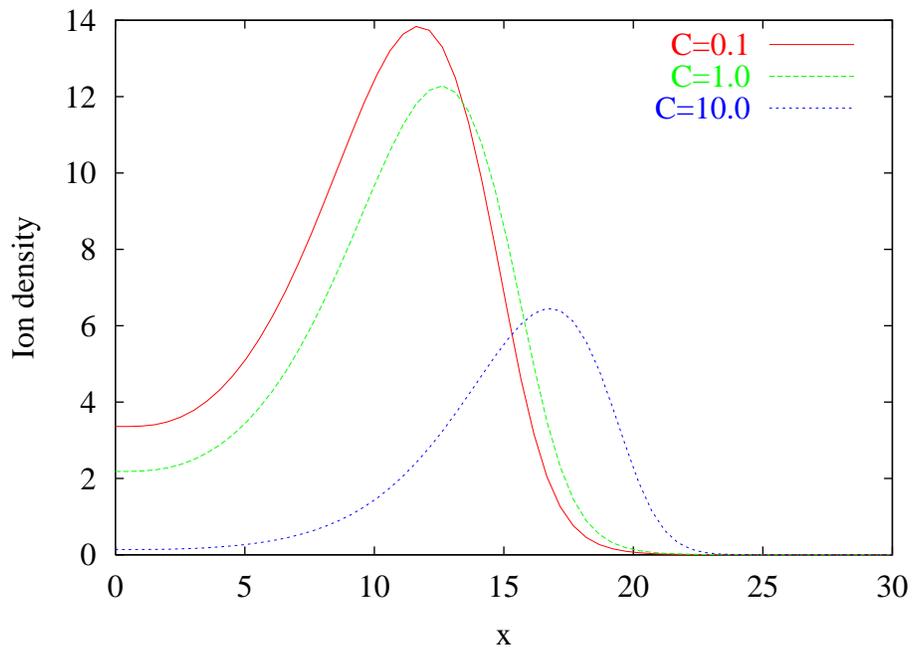} \par}

\caption{\label{Fig:ions}Ion density versus distance 
at the minimum \protect\( a_{C}\protect \) for different values of $C$.
}
\end{figure}

\end{document}